\documentclass{PoS}

\title{Dynamically generated resonances from vector meson-vector meson interaction based
on a hidden-gauge unitary approach}

\ShortTitle{Dynamically generated resonances from vector meson-vector meson interaction}

\author{\speaker{L. S. Geng}\thanks{lsgeng@ific.uv.es.}, E. Oset, R. Molina\\
Departamento de F\'{\i}sica Te\'orica and IFIC, Universidad de
Valencia-CSIC, E-46071 Valencia, Spain }

\author{D. Nicmorus\\
Fachbereich Theoretische Physik, Institut f\"ur Physik, Karl-Franzens-Universit\"at Graz, Universit\"atsplatz 5, A-8010 Graz, Austria}

\abstract{We present a recent study of vector meson-vector meson interaction using 
a unitary approach based on the hidden-gauge Lagrangians. We find that 11 states get dynamically generated, corresponding to poles of the scattering matrices on
the second Riemann sheet. Five of them are associated to the $f_0(1370)$, $f_0(1710)$, 
$f_2(1270)$, $f_2'(1525)$, and $K^*_2(1430)$, by comparing the mass, width, and decay pattern of
 these states with those of their experimental counterparts. Six others with the quantum numbers of
$h_1$, $a_0$, $b_1$, $a_2$, $K^*_0$, and $K_1$ can not be clearly identified with any experimentally
observed resonances and should be taken as predictions to be tested by future experiments.}

\FullConference{International Workshop on Effective Field Theories: from the pion to the upsilon \\
		February 2-6 2009\\
		Valencia, Spain\\}

\begin{document}
\section{Introduction}
The combination of coupled-channel unitarity and chiral Lagrangians, the so-called
unitary chiral approach, has been quite
successful in explaining the nature of and/or related data about some low-lying mesonic and baryonic excitations (see Ref.~\cite{Oller:2000ma} for an introduction and earlier references).  For instance, 
the scalar resonances $f_0(600)$, $f_0(980)$, and $a_0(980)$, appear
naturally in the unitarized Nambu-Goldstone-boson self-interactions~\cite{Oller:1997ti}; the 
$\Lambda(1405)$, on the other hand, is found originated from
the interaction between the baryon octet of the proton and the pseudoscalar octet of the pion~\cite{Kaiser:1995eg}. 
Because these states are dynamically generated from meson-meson or meson-baryon interactions,
they are viewed as composite particles or hadronic molecules, instead of genuine
$q\bar{q}$ or $qqq$ systems. Studies of their behavior in the large $N_c$ limit have
confirmed that they are largely meson-meson or meson-baryon
composite systems, though some of them may contain a small genuine $q\bar{q}$ or $qqq$ component~\cite{Pelaez:2003dy}.

Both chiral symmetry and coupled-channel unitarity play an important role in the success of
the unitary chiral approach. Chiral symmetry puts strong constraints on the allowed forms of the interaction Lagrangians. Coupled-channel unitarity, on the other hand,
extends the application region of chiral perturbation theory (ChPT). This is achieved in 
the Bethe-Salpeter equation method by summing over all the $s$-wave bubble diagrams.

Inspired by the success of the unitary chiral approach, a further extension has recently been taken to
study  the interaction between two vector mesons and between one vector meson and
one baryon~\cite{Molina:2008jw,Geng:2008gx,Molina:2009eb,Gonzalez:2008pv,Sarkar:2009kx}. The novelty is that instead of using interaction kernels provided by ChPT, one uses transition amplitudes provided by the hidden-gauge Lagrangians, which lead to
a suitable description of the interaction of vector mesons among themselves and of vector mesons with other mesons or baryons. Coupled-channel unitarity works in the same way as in the unitary chiral approach, but now the dynamics is provided by
the hidden-gauge Lagrangians~\cite{Bando:1984ej}.  As shown by several recent works~\cite{Molina:2008jw,Geng:2008gx,Molina:2009eb,Gonzalez:2008pv,Sarkar:2009kx}, this combination seems to work very well.

In this talk, we report on a recent study of vector meson-vector meson interaction using
the unitary approach 
where 11 states are found dynamically generated~\cite{Geng:2008gx}. First, we briefly explain the unitary approach. Then,  we show the main results and compare them with available data, followed by a short summary at the end.

\section{Theoretical framework}
In the following, we briefly 
outline the main ingredients of the unitary approach (details can be found in Refs.~\cite{Molina:2008jw,Geng:2008gx}). 
There are two basic building-blocks in this approach: transition amplitudes
provided by the hidden-gauge Lagrangians~\cite{Bando:1984ej} and a unitarization procedure. We adopt the Bethe-Salpeter equation method
 $T=(1-VG)^{-1}V$
 to unitarize the transition amplitudes $V$ for $s$-wave interactions, 
where $G$ is a diagonal matrix
of the vector meson-vector meson one-loop function
\begin{equation}
 G=i\int\frac{d^4q}{(2\pi)^4}\frac{1}{q^2-M_1^2}\frac{1}{q^2-M_2^2}
\end{equation}
with $M_1$ and $M_2$ the masses of the two vector mesons.

\begin{figure}[t]
\centerline{\includegraphics[scale=0.35]{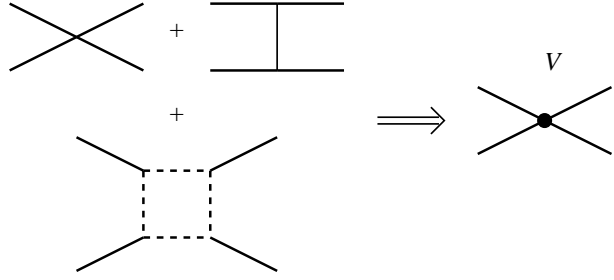}}
\caption{Transition amplitudes $V$ appearing in the coupled-channel Bethe-Salpeter equation.\label{fig:dia1}}
\end{figure}

In Refs.~\cite{Molina:2008jw,Geng:2008gx}
three mechanisms,  as shown in Fig.~\ref{fig:dia1}, have been taken into account for the calculation of the transition amplitudes $V$:
the four-vector-contact term, the t(u)-channel vector-exchange amplitude, and the direct box amplitude with two intermediate pseudoscalar mesons. Other possible mechanisms, e.g. crossed box amplitudes and
box amplitudes involving anomalous couplings, have been neglected, whose contributions are assumed to be small as has been explicitly shown to be the case for rho-rho scattering in Ref.~\cite{Molina:2008jw}.

Among the three mechanisms considered for $V$, the
four-vector-contact term and $t(u)$-channel vector-exchange one are responsible for
the formation of resonances or bound states if the interaction generated by them
is strong enough. In this sense,
the dynamically generated states can be thought of as ``vector meson-vector meson molecules.'' On the
other hand, the consideration of the imaginary part of the direct box amplitude allows the generated states to decay into two pseudoscalars. It should
be stressed that in the present approach these two mechanisms play quite different roles:
the four-vector-contact term and the $t(u)$-channel vector-exchange one are responsible
for generating the resonances whereas the direct box amplitude mainly contributes to their decays.

Since the energy regions we are interested in are close to the two vector-meson threshold, the three-momenta of the external vector mesons are smaller than the corresponding masses, $|\vec{q}|^2/M^2\ll1$, and therefore can be safely neglected. This considerably simplifies 
the calculation of the four-vector-contact term and the $t(u)$-channel vector-exchange amplitude, whose explicit expressions can be found in Appendix A of Ref.~\cite{Geng:2008gx}. To calculate the box diagram, one has
to further introduce two parameters, $\Lambda$ and $\Lambda_b$. The parameter $\Lambda$ regulates the four-point loop function, and $\Lambda_b$ is related to the form factors used for the vector-pseudoscalar-pseudoscalar vertex, which
is inspired by the empirical form factors used in the study of vector-meson decays~\cite{Titov:2000bn}.
In the present work we use $\Lambda=1$ GeV and $\Lambda_b=1.4$ GeV, unless otherwise stated.
The values of $\Lambda$ and
$\Lambda_b$ have been fixed in Ref.~\cite{Molina:2008jw} to obtain the widths of the $f_0(1370)$ and $f_2(1270)$. They are found to provide a good description of the widths of the $f'_2(1525)$, $K_2^*(1430)$, and $f_0(1710)$ as well (see Table I).

To regularize the two vector-meson one-loop function, Eq.~(2.1), one has to introduce either cutoffs in the cutoff method or subtraction constants in the dimensional regularization method. Further details concerning the values of these parameters can be found in Ref.~\cite{Geng:2008gx}. The results presented in this talk are obtained using the dimensional regularization method with the values of the subtraction constants given in Ref.~\cite{Geng:2008gx}. In particular, we have fine-tuned the values of three
subtraction constants to reproduce the masses of the three tensor states, i.e.,
the $f_2(1270)$, the $f'_2(1525)$, and the $K_2^*(1430)$. The masses of the other
eight states are predictions. 
\section{Results and discussions}
\begin{table*}[b]
      \renewcommand{\arraystretch}{1.4}
     \setlength{\tabcolsep}{0.1cm}
\caption{The properties, (mass, width) [in units of MeV], of the 11  dynamically
generated states and, if existing, of those of their PDG
counterparts~\cite{Amsler:2008zz}. The association of the dynamically generated states with their experimental counterparts is determined by matching their mass, width, and decay pattern.
\label{table:sum}}
\begin{center}
\begin{tabular}{c|c|cc|ccc}\hline\hline
$I^{G}(J^{PC})$&\multicolumn{3}{c|}{Theory} & \multicolumn{3}{c}{PDG data}\\\hline
              & Pole position &\multicolumn{2}{c|}{Real axis} & Name & Mass & Width  \\
              &               & $\Lambda_b=1.4$ GeV & $\Lambda_b=1.5$ GeV &           \\\hline
$0^+(0^{++})$ & (1512,51) & (1523,257) & (1517,396)& $f_0(1370)$ & 1200$\sim$1500 & 200$\sim$500\\
$0^+(0^{++})$ & (1726,28) & (1721,133) & (1717,151)& $f_0(1710)$ & $1724\pm7$ & $137\pm 8$\\
$0^-(1^{+-})$ & (1802,78) & \multicolumn{2}{c|} {(1802,49)}   & $h_1$\\
$0^+(2^{++})$ & (1275,2) & (1276,97) & (1275,111) & $f_2(1270)$ & $1275.1\pm1.2$ & $185.0^{+2.9}_{-2.4}$\\
$0^+(2^{++})$ & (1525,6) & (1525,45) &(1525,51) &$f_2'(1525)$ & $1525\pm5$ & $73^{+6}_{-5}$\\\hline
$1^-(0^{++})$    & (1780,133) & (1777,148) &(1777,172) & $a_0$\\
$1^+(1^{+-})$    & (1679,235) & \multicolumn{2}{c|}{(1703,188)} & $b_1$ \\
$1^-(2^{++})$    &  (1569,32) & (1567,47) & (1566,51)& $a_2(1700)??$
\\\hline
$1/2(0^+)$       &  (1643,47) & (1639,139) &(1637,162)&  $K_0^*$ \\
$1/2(1^+)$       & (1737,165) &  \multicolumn{2}{c|}{(1743,126)} & $K_1(1650)?$\\
$1/2(2^+)$       &  (1431,1) &(1431,56) & (1431,63) &$K_2^*(1430)$ & $1429\pm 1.4$ & $104\pm4$\\
 \hline\hline
    \end{tabular}
\end{center}
\end{table*}

Searching for poles of the scattering matrix $T$ on the second Riemann sheet, we find
11 states in nine strangeness-isospin-spin channels as shown in Table I. Theoretical masses and widths are obtained with two
different methods: In the ``pole position'' method, 
the mass corresponds to the
real part of the pole position on the complex plane and the width corresponds to twice
its imaginary part. In this case, the box diagrams
corresponding to decays into two pseudoscalars are not included.
In the "real axis" method, the resonance parameters are
obtained from the modulus squared of the amplitudes of the dominant channel of each state
 on the real axis\footnote{See Tables I, II, and III of Ref.~\cite{Geng:2008gx}.}, where the mass corresponds to the energy at which the modulus squared has a maximum and the width corresponds to the difference
between the two energies where the modulus squared is half of the
maximum value. In this latter case, the box amplitudes are included.  The results
shown in Table I have been obtained using two different values of $\Lambda_b$,
which serve to quantify the uncertainties related to this parameter.

Our treatment of the box amplitudes enables us to obtain
the decay branching ratios of the generated states into two pseudoscalar mesons using the real-axis method in the
following way:
\begin{itemize}
 \item First,  we calculate the width of the selected state with and without the contributions of
the box diagrams,
 $\Gamma(\mathrm{total})=\Gamma(\mathrm{VV})+\Gamma(\mathrm{PP})$ and $\Gamma(\mathrm{VV})$.

\item Second,  we estimate the partial decay width into a particular two-pseudoscalar channel of the state by including only the contribution of the particular channel. Taking the $\pi\pi$ channel as an example, this way we obtain $\Gamma(\mathrm{w}\pi\pi)$.
The contribution of the $\pi\pi$ channel is then determined as $\Gamma({\pi\pi})=\Gamma(\mathrm{w}\pi\pi)-\Gamma(\mathrm{VV})$ and its branching ratio is calculated
as $\Gamma({\pi\pi})/\Gamma(\mathrm{total})$. The partial decay branching ratios into other two-pseudoscalar channels are calculated similarly. It should be noted that we have assumed
that interference between contributions of different channels is small, which seems
to be justified since the sum of the calculated partial decay widths agrees well with the total decay width. 
\end{itemize}
The so-obtained branching ratios are given in Tables II, III, and IV, in comparison with available data~\cite{Amsler:2008zz}. 

From Table II, one can see that our results for the
two $f_2$ states agree very well with the data. For the $f_0(1370)$, according to the PDG~\cite{Amsler:2008zz},
the $\rho\rho$ mode is dominant. 
In our approach, however, the $\pi\pi$ mode is dominant, which is consistent with the results of 
Ref.~\cite{Albaladejo:2008qa}
and the recent analysis of D. V. Bugg~\cite{Bugg:2007ja}. For the $f_0(1710)$, using the branching
ratios given in Table I, we obtain $\Gamma(\pi\pi)/\Gamma(K\bar{K})<1\%$ and $\Gamma(\eta\eta)/\Gamma(K\bar{K})\sim49\%$.
On the other hand, the PDG gives the following
averages: $\Gamma(\pi\pi)/ \Gamma(K\bar{K}) =0.41_{-0.17}^{+ 0.11}$,
and $\Gamma(\eta\eta)/\Gamma(K \bar{K})=0.48 \pm0.15$~\cite{Amsler:2008zz}.  Our calculated branching ratio
for the $\eta\eta$ channel is in agreement with their average, while the ratio for
the $\pi\pi$ channel is much smaller. However, we notice that
the above PDG $\Gamma(\pi\pi)/ \Gamma(K\bar{K})$ ratio is taken from the BES data on
$J/\psi\rightarrow \gamma\pi^+\pi^-$~\cite{Ablikim:2006db}, which comes from
a partial wave analysis that includes seven resonances. On the other hand,
the BES data on $J/\psi\rightarrow\omega K^+K^-$~\cite{Ablikim:2004st} give an upper limit
$\Gamma(\pi\pi)/ \Gamma(K\bar{K})<11\%$ at the $95\%$ confidence level. Clearly more analysis is
advised to settle the issue.

Compared to the decay branching ratios into 
two pseudoscalar mesons of the isospin 0 states, those of the isospin 1 states with spin either 0 or 2 are relatively small, as shown in Table III.

In Table IV, one can see that the dominant decay mode of the $K^*_2(1430)$ is $K\pi$ both theoretically and experimentally. However, other modes, such as $\rho K$, $K^*\pi$, and $K^*\pi\pi$,
account for half of its decay width according to the PDG~\cite{Amsler:2008zz}. This is consistent with the fact that our $K^*_2(1430)$ is narrower than its experimental counterpart, as can be seen from
Table I.

 The three spin 1 states with the quantum numbers of $h_1$, $b_1$ and $K_1$, do not decay into two pseudoscalars in our approach since
the box diagrams do not contribute as explained in Ref.~\cite{Geng:2008gx}.

\begin{table}
      \renewcommand{\arraystretch}{1.4}
     \setlength{\tabcolsep}{0.2cm}
\caption{Branching ratios of the $f_0(1710)$, $f_0(1370)$, $f_2(1270)$, and $f'_2(1525)$ in comparison with data~\cite{Amsler:2008zz}.}
\begin{tabular}{c|cc|cc|cc|cc}
\hline\hline
 &\multicolumn{2}{c|}{$\Gamma(\pi\pi)/\Gamma(\mathrm{total})$}&\multicolumn{2}{c|}{$\Gamma(\eta\eta)/\Gamma(\mathrm{total})$}&\multicolumn{2}{c|}{$\Gamma(K\bar{K})/\Gamma(\mathrm{total})$}
&\multicolumn{2}{c}{$\Gamma(\mathrm{VV})/\Gamma(\mathrm{total})$}\\
& Our model & Data & Our model & Data & Our model & Data & Our model & Data\\
\hline 
$f_0(1370)$ & $\sim72\%$ & & $<1\%$ & & $\sim10\%$ & &$\sim18\%$ &\\
$f_0(1710)$ & $<1\%$    &  & $\sim27\%$ &  & $\sim55\%$&& $\sim18\%$ &\\
$f_2(1270)$ & $\sim88\%$ & $84.8\%$ & $ <1\%$ & $<1\%$ & $\sim 10\%$&$4.6\%$ & $<1\%$ &\\
$f'_2(1525)$ &$<1\%$ & $0.8\%$ & $\sim21\%$ & $10.4\%$ & $\sim 66\%$& $88.7\%$ & $\sim13\%$&\\
 \hline\hline
\end{tabular}
\end{table}
\begin{table}
\begin{center}
      \renewcommand{\arraystretch}{1.4}
     \setlength{\tabcolsep}{0.3cm}
\caption{Branching ratios of the $a_0$ and $a_2$ states.}
\begin{tabular}{c|ccc}
\hline\hline
 &$\Gamma(K\bar{K})/\Gamma(\mathrm{total})$&$\Gamma(\pi\eta)/\Gamma(\mathrm{total})$&$\Gamma(\mathrm{VV})/\Gamma(\mathrm{total})$\\
\hline 
$a_0$ & $\sim27\%$ &  $\sim23\%$ &  $\sim50\%$ \\
$a_2$ & $\sim21\%$ & $\sim17\%$ & $\sim62\%$\\
 \hline\hline
\end{tabular}
\end{center}
\end{table}

\begin{table}
      \renewcommand{\arraystretch}{1.4}
     \setlength{\tabcolsep}{0.3cm}
\begin{center}
\caption{Branching ratios of the $K^*_0$ and $K^*_2(1430)$ states in comparison with data~\cite{Amsler:2008zz}.}
\begin{tabular}{c|cc|cc|cc}
\hline\hline
 &\multicolumn{2}{c|}{$\Gamma(K\pi)/\Gamma(\mathrm{total})$}&\multicolumn{2}{c|}{$\Gamma(K\eta)/\Gamma(\mathrm{total})$}&\multicolumn{2}{c}{$\Gamma(\mathrm{VV})/\Gamma(\mathrm{total})$}\\
& Our model & Data & Our model & Data & Our model & Data \\
\hline 
$K^*_0$ & $\sim65\%$ & & $\sim9\%$ & & $\sim26\%$ \\
$K^*_2(1430)$ & $\sim93\%$ & $49.9\%$  & $\sim5\%$ & $<1\%$ & $\sim2\%$ \\
 \hline\hline
\end{tabular}
\end{center}
\end{table}
It is interesting to note that out of the 21 combinations of
strangeness, isospin and spin, we have found resonances only in nine of
them. In all the ``exotic'' channels, from the point of view that they cannot be formed from $q\bar{q}$
combinations, we do not find dynamically
generated resonances, including the three (strangeness=0, isospin=2)
channels, the three (strangeness=1, isospin=3/2) channels, the six
strangeness=2 channels (with either isospin=0 or isospin=1).
On the other hand, there do exist some structures on the real axis.
For instance, in the (strangeness=0, isospin=2) channel, one finds a
dip around $\sqrt{s}=1300$ MeV in the spin=0 channel, and a broad
bump in the spin=2 channel around $\sqrt{s}=1400$ MeV, as can be
clearly seen from Fig.~7 of Ref.~\cite{Geng:2008gx}. In the
(strangeness=1, isospin=3/2) and (strangeness=2, isospin=1) channels,
one observes similar structures occurring at shifted energies due to
the different masses of the $\rho$ and the $K^*$, as can be seen
from Figs.~8 and 9 of Ref.~\cite{Geng:2008gx}. However, these states do not correspond to poles on the complex plane, and hence, according to the common criteria, they do
not qualify as resonances.

\section{Summary and conclusions}
We have performed a study of vector meson-vector meson interaction
using a unitary approach. Employing the coupled-channel
Bethe-Salpeter equation to unitarize the transition
amplitudes provided by the hidden-gauge Lagrangians, we find that 11 states get dynamically generated
in nine strangeness-isospin-spin channels. Among them, five states are associated to those reported in
the PDG, i.e., the $f_0(1370)$, the $f_0(1710)$, the $f_2(1270)$,
the $f'_2(1525)$, the $K_2^*(1430)$. The association of two other
states, the $a_2(1700)$ and the $K_1(1650)$, are likely, particularly the
$K_1(1650)$, but less
certain. Four of the 11 dynamically generated states can not be
associated with any known states in the PDG. Another interesting finding of our work is that the broad bumps found in
four exotic channels  do not correspond to poles on the
complex plane and, hence, do
not qualify as resonances.

\section{Acknowledgments}
L. S. Geng thanks R. Molina,  L. Alvarez-Ruso, and M. J. Vicente Vacas for
useful discussions. This work is partly
supported by DGICYT Contract No. FIS2006-03438 and the EU Integrated
Infrastructure Initiative Hadron Physics Project under contract
RII3-CT-2004-506078.  L.S.G. acknowledges support from the MICINN in the Program ``Juan de la Cierva.''

\end{document}